\documentclass[aps,prl,twocolumn,showpacs,preprintnumbers,nofootinbib]{revtex4}

\usepackage{graphicx}
\usepackage[FIGTOPCAP]{subfigure}
\usepackage{dcolumn}
\usepackage{bm}
\usepackage[normalem]{ulem}

\usepackage{amsmath}
\usepackage{epstopdf}
\usepackage{amsmath,amssymb}

\newcommand{\beq}{\begin{eqnarray}}
\newcommand{\eeq}{\end{eqnarray}}

\newcommand{\real}{{\sf I}\kern-.12em{\sf R}}
\newcommand{\comp}{{\sf I}\kern-.50em{\sf C}}
\newcommand{\unity}{{\sf I}\kern-.54em{\sf 1}}

\newcommand{\tla}{\tilde\lambda}

\begin{document}

\title{
Thermal monopole condensation in QCD with physical quark masses}

\author{Marco Cardinali}
\email{marco.cardinali@pi.infn.it}
\affiliation{Dipartimento di Fisica dell'Universit\`a di Pisa, Largo Pontecorvo 3, I-56127 Pisa, Italy}
\affiliation{INFN - Sezione di Pisa, Largo Pontecorvo 3, I-56127 Pisa, Italy}

\author{Massimo D'Elia}
\email{massimo.delia@unipi.it}
\affiliation{Dipartimento di Fisica dell'Universit\`a di Pisa, Largo Pontecorvo 3, I-56127 Pisa, Italy}
\affiliation{INFN - Sezione di Pisa, Largo Pontecorvo 3, I-56127 Pisa, Italy}

\author{Andrea Pasqui}
\email{andreP7@hotmail.it}
\affiliation{Dipartimento di Fisica dell'Universit\`a di Pisa, Largo Pontecorvo 3, I-56127 Pisa, Italy}
\affiliation{INFN - Sezione di Pisa, Largo Pontecorvo 3, I-56127 Pisa, Italy}

\date{\today}

\begin{abstract}
Thermal monopoles, identified after Abelian projection
as magnetic currents wrapping non-trivially around the thermal circle,
are studied in $N_f = 2+1$ QCD at the physical point.
The distribution in the number of wrappings, which in pure gauge
theories points to a condensation temperature coinciding with 
deconfinement, points in this case to around 275 MeV,
almost twice the QCD crossover temperature $T_c$;
similar indications emerge looking for the formation 
of a percolating current cluster. The 
possible relation with other non-perturbative phenomena
observed above $T_c$ is discussed.
\end{abstract}


\maketitle

Color confinement is one of the most intriguing aspects of 
Quantum Chromo Dynamics (QCD), the theory of strong interactions.
Despite a plenty of phenomenological and numerical evidence, a full
theoretical understanding of it, stemming from QCD first principles,
is still lacking. In pure $SU(N)$ gauge theories, confinement 
can be reconduced to the realization of an exact center symmetry: when 
this is not spontaneously broken, one has a linearly rising 
potential between static color charges, with an associated string
tension. Exact order parameters, 
like the Polyakov loop, can then be easily associated with 
the high $T$ deconfinement transition.
As one moves to full QCD, 
center symmetry gets broken explicitly by dynamical quarks 
and the confining string breaks at large distances.
The restoration of chiral symmetry, which is almost exact, becomes the 
dominant phenomenon which identifies a (pseudo)critical crossover temperature
around 155~MeV~\cite{aefks,afks,betal,tchot,tchot2}.
While the Polyakov loop and its susceptibilities
still show a non-trivial behaviour around a similar temperature~\cite{Petreczky:2020olb}, a clear association with deconfinement is not clear.

On the other hand, various mechanisms have been
proposed, which tipically interpret confinement in terms of the 
condensation of topological degrees of freedom. 
Even if no consensus yet exists regarding the nature
of such degrees of freedom,
all descriptions lead to a correct identification of 
the deconfinement transition in pure gauge theories.
It is therefore of great interest to investigate them 
in full QCD, where a univocal description of deconfinement is missing.
A possible mechanism 
is that based on dual superconductivity~\cite{thooft78, Mandelstam:1974pi},
i.e. on the idea that the QCD vacuum is characterized by the
spontaneous breaking of an Abelian magnetic symmetry, 
induced by the condensation of magnetic charges, 
leading to confinement of chromo-electric charges via a dual Meissner effect.
The mechanism has been tested by lattice simulations
in various ways, like looking at 
the expectation value of magnetically charged operators
and at the effective monopole 
action~\cite{DelDebbio:1995yf, DiGiacomo:1999yas, DiGiacomo:1999fb, Carmona:2001ja, Carmona:2002ty, DElia:2005sfk, Bonati:2011jv, Chernodub:1996ps,bari,DAlessandro:2006hfn,DiGiacomo:2020fzr}, or by studying the properties of monopole currents extracted
from non-Abelian gauge configurations.
The identification of Abelian degrees of freedom 
relies on a 
procedure known as Abelian projection, which is based 
on the choice of an adjoint
field. No natural adjoint field exists in QCD, 
so that the procedure is partially arbitrary: a popular choice is 
the so-called Maximal Abelian gauge (MAG) projection.

A particularly useful way to look for monopole condensation
is to start from the deconfined, high-temperature phase and 
to investigate the properties of the so-called thermal monopoles, 
which are quasi-particles identified with  
monopole currents with a non-trivial wrapping around 
the Euclidean time direction~\cite{Chernodub:2006gu, Chernodub:2007cs, Bornyacovmm, Ejiri:1995gd, shuryak,liao,ratti}.  
An analogy with the path-integral formulation for a system of identical
particles permits to associate monopole currents 
with multiple wrappings around the thermal circle
as set of thermal monopoles undergoing a permutation 
cycle~\cite{cristof}:
based on this analogy, the statistical distribution 
of the multiple wrappings permits to reconstruct 
the quantum properties of the thermal monopole ensemble and to 
extrapolate a temperature where a phenomenon 
similar to Bose-Einstein condensation (BEC) takes place.

Notably, this approach returns condensation temperatures which coincide,
within errors and both for $SU(2)$ and $SU(3)$ pure gauge 
theories~\cite{DAlessandro:2010jdd, Bonati:2013bga},
with the standard deconfinement temperature;
that happens in a peculiar way
 also for trace deformed theories~\cite{Bonati:2020lal} where, in spite 
of the decrease of the thermal monopole density approaching the 
center symmetric phase, monopole condensation, revealed by the 
distribution in the number of thermal wrappings, takes place where 
expected anyway.
Based on these successes, the plan of the present study
is to extend the investigation to QCD with $N_f = 2+1$ flavors at
the physical point,
to understand if and where a BEC-like phenomenon takes place for thermal
monopoles in this case.

{\it Technical details} -- 
Abelian monopoles in $SU(N)$ gauge theories are identified after the so-called Abelian projection \cite{tHooft:1981bkw}. 
The $SU(N)$ gauge symmetry is fixed apart from a remnant $U(1)^{N-1}$ Abelian symmetry:
in absence of a natural Higgs field in the theory, the way this is done is to 
some extent arbitrary.
We consider here the so-called generalized MAG~\cite{Stack:2001hm}, 
following the implementation of Ref.~\cite{Bonati:2013bga}.
On the lattice, the gauge is fixed by maximizing 
\begin{equation}
\tilde F_{\rm MAG} = \sum_{\mu,n} 
\mbox{tr} \left( U_{n; \mu} \tilde\lambda U^{\dagger}_{n ; \mu} \, \tilde\lambda \right) 
\label{magsu3_2}
\end{equation}
where $\tilde \lambda$ is a generic element of the Cartan subalgebra
and $U_{n; \mu}$ is the gauge link variable in position $n$ and direction $\mu$.
For appropriate choices of $\tla$, such maximization makes diagonal 
the following operator
\begin{equation}
\tilde X(n) = \sum_\mu \left[ U_{n; \mu} \tilde\lambda U^\dagger_{n ; \mu}
+U^\dagger_{n-\hat\mu; \mu} \tilde\lambda U_{n-\hat\mu; \mu} \right]
\label{XdefN}
\end{equation}
which plays the role of the Higgs field. 
If $\tla$ is expanded over the basis of fundamental weights $\phi_0^k$ ($k = 1, N-1$)
\beq
\tla = 
b^k \phi_0^k , \ \
\label{tla1}
\phi_{0}^{k}=\frac{1}{N}\, {\rm diag}\, (\, \underbrace{N\hspace{-3pt}-\hspace{-3pt}k,\ldots N\hspace{-3pt}-\hspace{-3pt}k}_{k}\, , 
\underbrace{ -\hspace{-1pt} k ,\,\ldots  -\hspace{-2pt} k }_{N-k})
\eeq 
the condition is that $b^k \neq 0\ \forall\ k$~\cite{Bonati:2013bga},
morover if they are all positive one has a well defined ordering 
for the Higgs field eigenvalues.
In this gauge (unitary gauge) and in the continuum, the Abelian 't Hooft field strength tensor~\cite{thooft74}
$F^{(k)}_{\mu\nu}$
can be recontructed, for each $U(1)$ residual subgroup, in terms of the
Abelian gauge field~\cite{DelDebbio:2002nb, DiGiacomo:2008gf} 
\begin{equation}
a^{(k)}_\mu \equiv {\rm tr} (\phi_0^k A_\mu) = \sum_{j = 1}^k ({A_\mu})_{jj} \, .
\label{amudef}
\end{equation}
Magnetic monopoles are 
identified as defect lines (in 4D) where two adiacent eigenvalues of $\tilde X$
coincide; one has $N-1$ independent monopole
species in correspondence of the residual $U(1)^{N-1}$ Abelian symmetry.
Instead in standard MAG~\cite{Brandstater:1991sn}, where one simply maximizes the diagonal part 
of gauge link variables, the residual gauge group includes 
a permutation symmetry which mixes different monopole species and makes 
them not well defined~\cite{Bonati:2013bga}.
As in Ref.~\cite{Bonati:2013bga} we consider a choice of $\tla$ 
with equal coefficients $b_k = 1$,
namely $\tla = {\rm diag} (1,0,-1)$ for $SU(3)$, 
so that all species are treated equally, leading also to 
a reduction of lattice artefacts~\cite{Bonati:2013bga}.

On the lattice, Eq.~(\ref{amudef}) is implemented
by taking the phases 
${\rm diag} (\phi^1_\mu(n),\phi^2_\mu(n), \dots \phi^N_\mu(n))$
of the diagonal part of gauge links $U_{n;\mu}$, then the 
$k$-th 't Hooft tensor $\theta _{\mu\nu}^{(k)}$ (Abelian plaquette) is built
in terms of the Abelian gauge phases 
\begin{equation}
\theta^k_\mu(n) = \sum_{j = 1}^{k} \phi^j_\mu(n) \, .
\label{thetadef}
\end{equation}
Singular points where eigenvalues coincide are not well defined on the lattice,
so monopole currents $m_{\mu}^{(k)}$ are detected as violations of the Bianchi identity, 
i.e.~by measuring magnetic fluxes across 
the elementary 3-cubes (DeGrand-Touissant construction~\cite{DeGrand:1980eq}):
\begin{equation}
	m_{\mu}^{(k)} = \frac{1}{12}\varepsilon _{\mu\nu\rho	\sigma} \hat{\partial}_{\nu}\bar{\theta}_{\rho\sigma}^{(k)}, 
\label{eq:monopole_current}
\end{equation}
where $\hat{\partial}_{\nu}$ is the forward lattice derivative and 
\begin{equation}
	\theta _{\mu\nu}^{(k)} \equiv \bar{\theta _{\mu\nu}^{(k)}} + 2\pi n_{\mu\nu}^{(k)}, \ \ \bar{\theta _{\mu\nu}^{(k)}} \in [0, 2\pi ), \ \ n_{\mu\nu}^{(k)} \in \mathbb{Z}.
\label{eq:condition_abelian_plaq}
\end{equation}
The current $m_{\mu}^{(i)}$ forms closed loops, i.e. $\partial _{\mu} m_{\mu}^{(i)} = 0$. Then
thermal monopoles (anti-monopoles) are identified with magnetic currents having a non trivial wrapping around the temporal direction. This thermal monopoles trajectories can be interpreted as the path-integral representation of an ensemble of magnetically charged quasi-particles populating the thermal medium \cite{Chernodub:2006gu, Chernodub:2007cs, Ejiri:1995gd, DAlessandro:2007lae}. In this picture, trajectories wrapping $k$ times around the thermal direction are associated with the cyclic permutation of $k$ identical quasi-particles~\cite{DAlessandro:2010jdd, Bonati:2013bga}. 
We will consider the total thermal monopoles density 
$\rho$ and its cycle decomposition $\rho_k$ defined as follows:
\begin{equation}
	\rho = \sum _{k} k \rho _k \, ; \ \ \ \rho_k \equiv \frac{N_{\mathrm{wrap}, k}}{V_s},
\label{eq:mon_density}
\end{equation}
where $V_s=a^3 L^3$ is the spatial volume and $N_{\mathrm{wrap}, k}$ is the number of 
currents wrapping $k$ times. 
The distribution in the number of wrappings $k$ can signal the approach to a BEC transition.
In particular
one expects
\begin{equation}
	\rho (k) \propto \frac{\mathrm{e}^{-\hat{\mu}/k}}{k^{\alpha}},
\label{eq:mon_condensation}
\end{equation}
where $\hat{\mu} \equiv - \mu /T$ is the dimensionless chemical potential and $\alpha = 5/2$ for non-interacting bosons. Monopole condensation is then signalled by $\hat{\mu}$ approaching zero
as a function of $T$~\cite{DAlessandro:2010jdd, Bonati:2013bga}.

\begin{figure}[t!]
\begin{center}
\includegraphics*[width=0.47\textwidth]{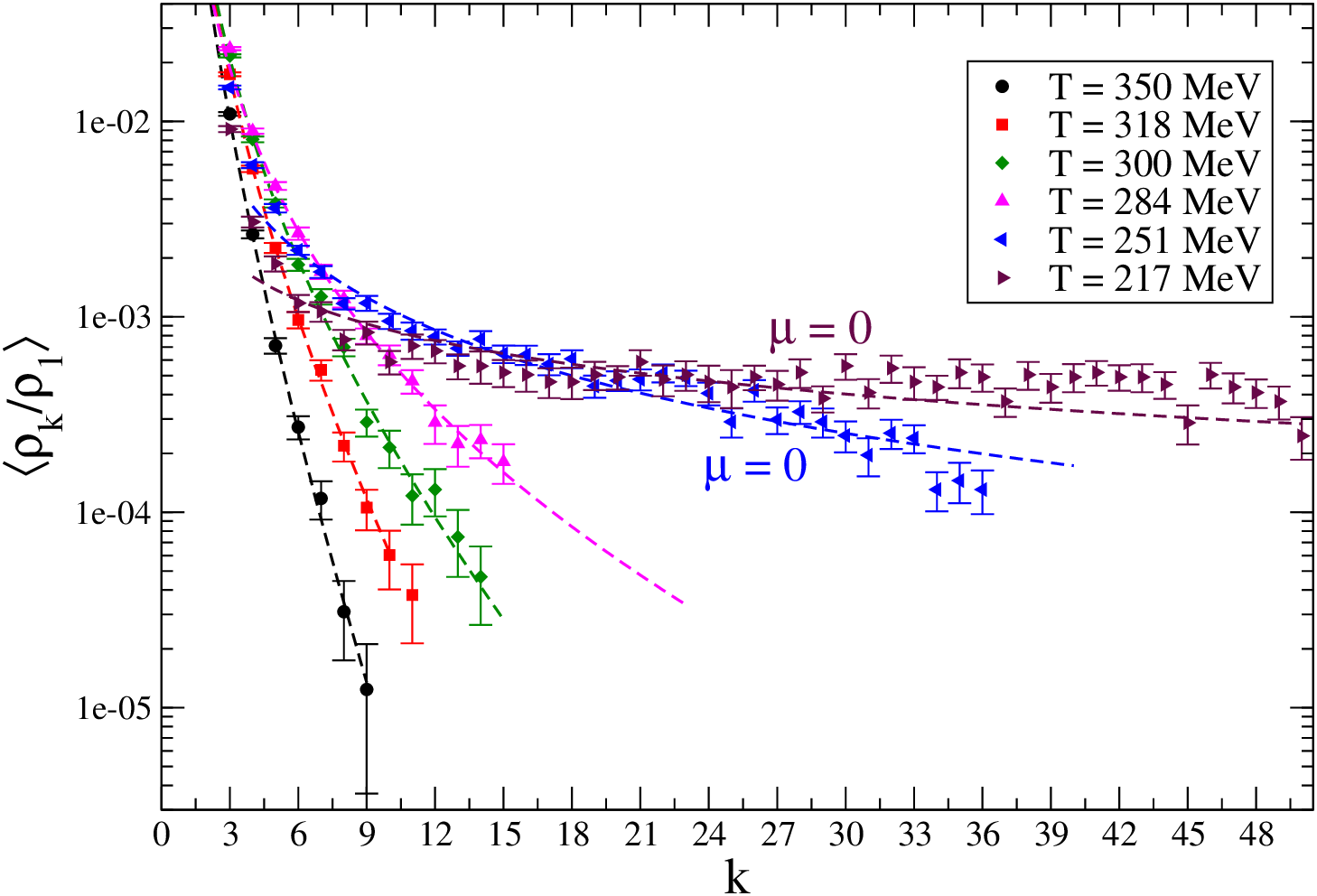}
\end{center}
\caption{Results for 
$\rho_k / \rho_1$ from simulations on the $48^3 \times 6$ lattice. The dashed
lines correspond to best fits to Eq.~(\ref{eq:mon_condensation}),
respectively fixing $\alpha = 5/2$ (for $T > 280$~MeV) 
or $\hat\mu = 0$ (for $T < 280$~MeV, the fit returns $\alpha \sim 1$ in this case). 
}
\label{fig1}
\end{figure}

Our discretization of $N_f = 2+1$ QCD relies on stout rooted staggered fermions 
and the Symanzik tree-level improved gauge action,
the partition function reads:
\begin{equation}
	Z = \int \mathcal{D}Ue^{-S_{YM}}\prod _{u, s, d} \mathrm{det} \left( D^f_{st}\right) ^{1/4},
\label{eq:action_disc}
\end{equation}
where $\mathcal{D}U$ is the product of Haar measures for $SU(3)$ gauge links. The gauge action $S_{YM}$ reads:
\begin{equation}
	S_{YM} = -\frac{\beta}{3}\sum _{i, \mu \neq \nu} \left( \frac{5}{6}P_{i, \mu\nu}^{1\times 1} -\frac{1}{12}P_{i, \mu\nu}^{2\times 1} \right).
\label{eq:gauge_action}
\end{equation}
where $P_{i, \mu\nu}^{1\times 1}$ and $P_{i, \mu\nu}^{2\times 1}$  are the real part of the trace of $1\times 1$ and $1\times 2$ loops. The staggered Dirac operator is:
\begin{equation}
	{D^f_{st}}_{\, ij} \hspace{-2pt} 
=  am_f\delta _{ij} + \sum _{\nu = 1}^{4} \frac{\eta _{i;\nu}}{2} (U^{(2)}_{i;\nu}\delta _{i; j-\hat{\nu}} 
	- U^{(2) \dagger}_{i-\hat{\nu}; \nu}\delta _{i, j +\hat{\nu}}) \hspace{-2pt} 
\label{eq:dirac_matrix}
\end{equation}
where $\eta _{i;\nu}$s are the staggered phases and $U^{(2) \dagger}_{i; \mu}$ is the two-time stout smeared link with isotropic smearing parameter $\rho = 0.15$.
 The bare gauge coupling $\beta$ and quark masses were kept on a line of constant physics \cite{Aoki:2009sc, Borsanyi:2010cj, Borsanyi:2013bia}
corresponding to a physical spectrum. Gauge configurations were generated using a 
Rational Hybrid Monte-Carlo algorithm running on GPUs~\cite{Bonati:2018wqj}.
We explored temperatures up to $\sim 1$ GeV and, in order to estimate finite spacing 
and finite size effects, we considered simulations on $32^3 \times 8$, $24^3 \times 6$ and 
$48^3 \times 6$ lattices.
Gauge fixing, namely the maximization of Eq.~(\ref{magsu3_2}), was based on an over-relaxation algorithm
(see Ref.~\cite{Bonati:2013bga} for more details). We report results
only for the first monopole species, the other coinciding within errors; 
different species correlations will be discussed in a forthcoming publication.

{\it Numerical Results} - The most striking aspect of our results emerges already from Fig.~\ref{fig1}, where we report 
the ratio $\rho_k / \rho_1$ as a function of $k$ for various temperatures. While results
for $T > 280$~MeV show a clear exponential decay 
with $k$ and can be nicely fitted to Eq.~(\ref{eq:mon_condensation}) with a non-zero
value of $\hat \mu$, for lower temperatures the dependence on $k$ is much flatter and 
actually compatible with $\hat \mu = 0$, as if monopole condensation were already at work.

\begin{figure}[t!]
\includegraphics*[width=0.45\textwidth]{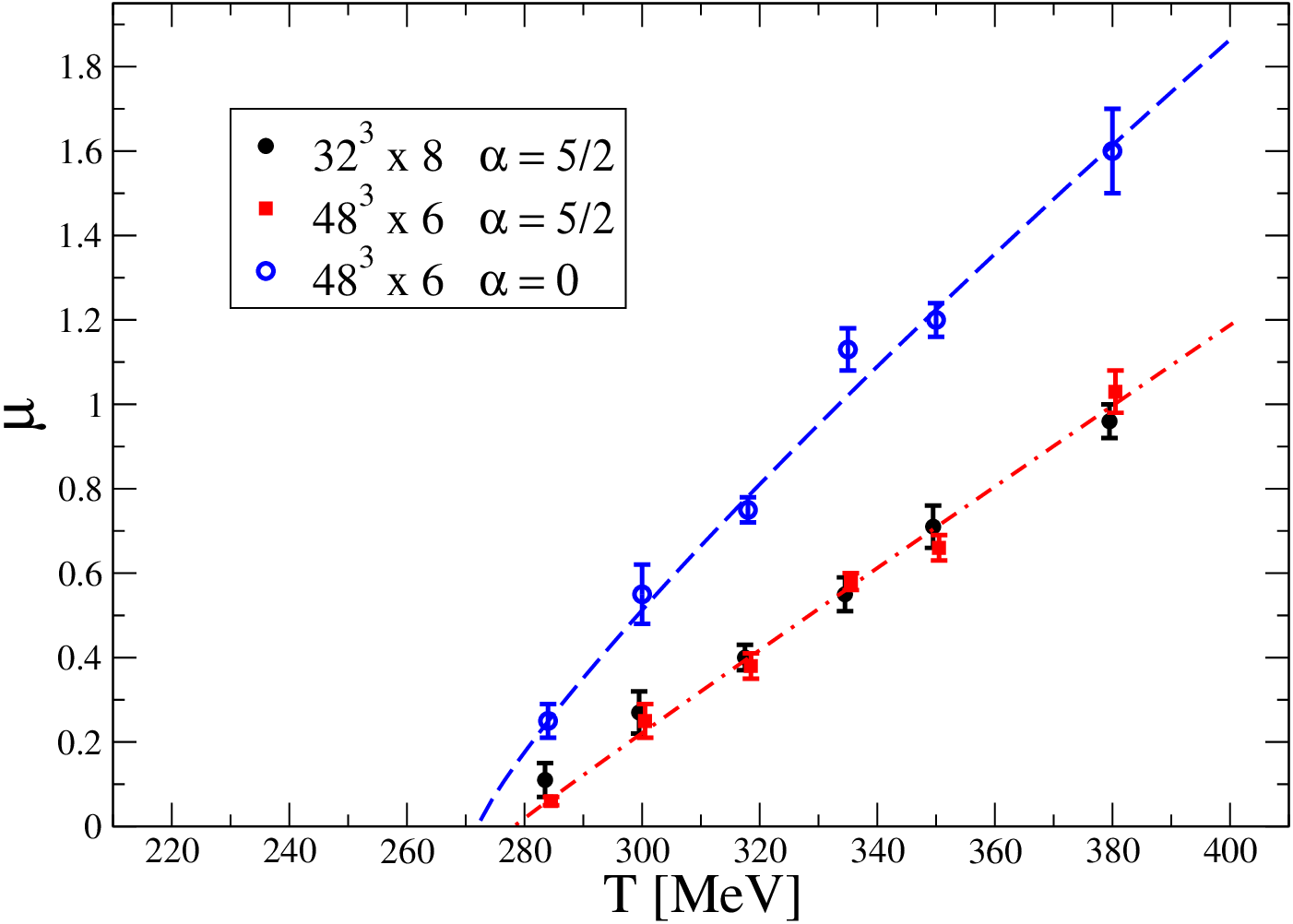}
\caption{Results for $\hat \mu$ at $T > 280$~MeV. The dashed
lines represent best fits of the $48^3 \times 6$ data to 
a critical behavior $\hat \mu (T) = a(T- T_{BEC})^{b}$,
returning $T_{BEC} = 272(2)$ ($b \sim 0.9$) and 
$T_{BEC} = 278(6)$ ($b \sim 1$) respectively
for $\alpha = 0$ and $\alpha = 5/2$. 
}
\label{fig2}
\end{figure}

The fitted values of $\hat \mu$ are displayed in Fig.~\ref{fig2}.
We report values obtained both for $\alpha = 0$ and $\alpha = 5/2$, 
showing that, as for pure gauge, the outcome is independent
of the assumption for it: in both cases $\hat \mu$ approaches 
zero at $T_{BEC} \sim 275$~MeV.
One can also appreciate that the dependence on the lattice spacing is 
negligible.
This result is not easy to interpret: 
$T_{BEC}$ is almost twice the well established
pseudocritical temperature of QCD, $T_c \simeq 155$~MeV,
at which chiral symmetry is restored.

Since monopole condensation
has been investigated in the literature in various different ways,
we decided, to confirm this strange result, to explore an alternative method, looking for 
the formation of a dominating cluster of monopole currents \cite{Ivanenko:1991wt}.
In practice, for each gauge configuration, we divide the whole set of monopole currents
$m_\mu$ of a given species into subsets (clusters) of connected currents and 
measure the ratio $r_c$ of the current length of the biggest cluster to the total
length of the whole set. In general $r_c \in [0,1]$ and one expects 
$\langle r_c \rangle \to 0$ in the thermodynamical limit if no dominating cluster 
forms, while it tends to some non-zero value when the largest cluster percolates
becoming microscopic. 

\begin{figure}
\begin{center}
\includegraphics*[width=0.45\textwidth]{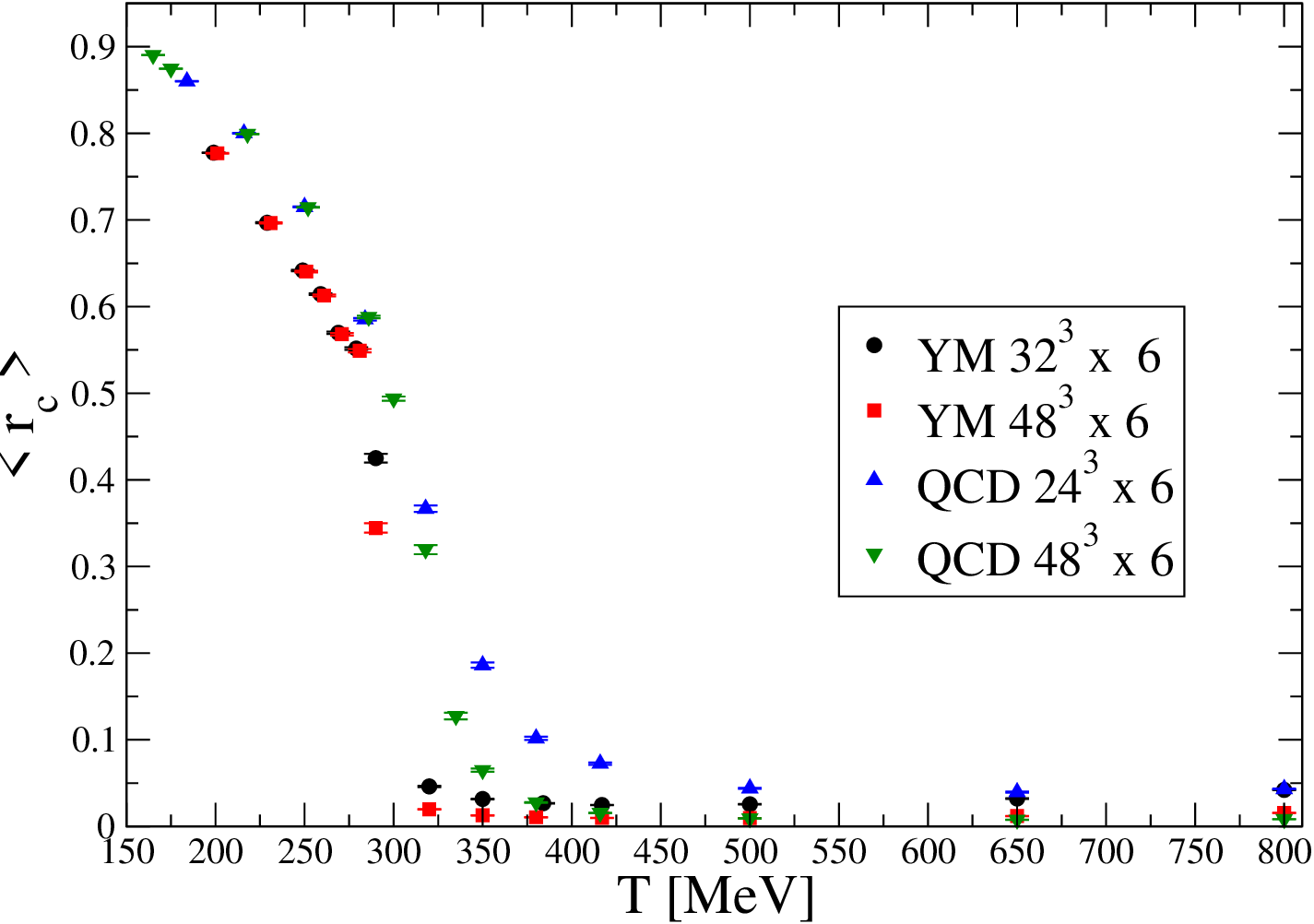}
\end{center}
\caption{Relative weight of the largest monopole current cluster as a function
of $T$, for full QCD and for the pure gauge theory.}
\label{fig3}
\end{figure}

In Fig.~\ref{fig3} we report data for $\langle r_c \rangle$ as a function of $T$
for two spatial volumes, comparing them with the same quantity computed
in pure gauge $SU(3)$.
The behavior is strikingly similar, with $\langle r_c \rangle$ becoming 
volume independent and approaching 1 when $T < 300$~MeV: the transition is just
sharper for pure gauge $SU(3)$, where a weak first order deconfining transition 
takes places at $T_{c,SU3} \sim 290$~MeV. 

\begin{figure}[t!]
\begin{center}
\includegraphics*[width=0.45\textwidth]{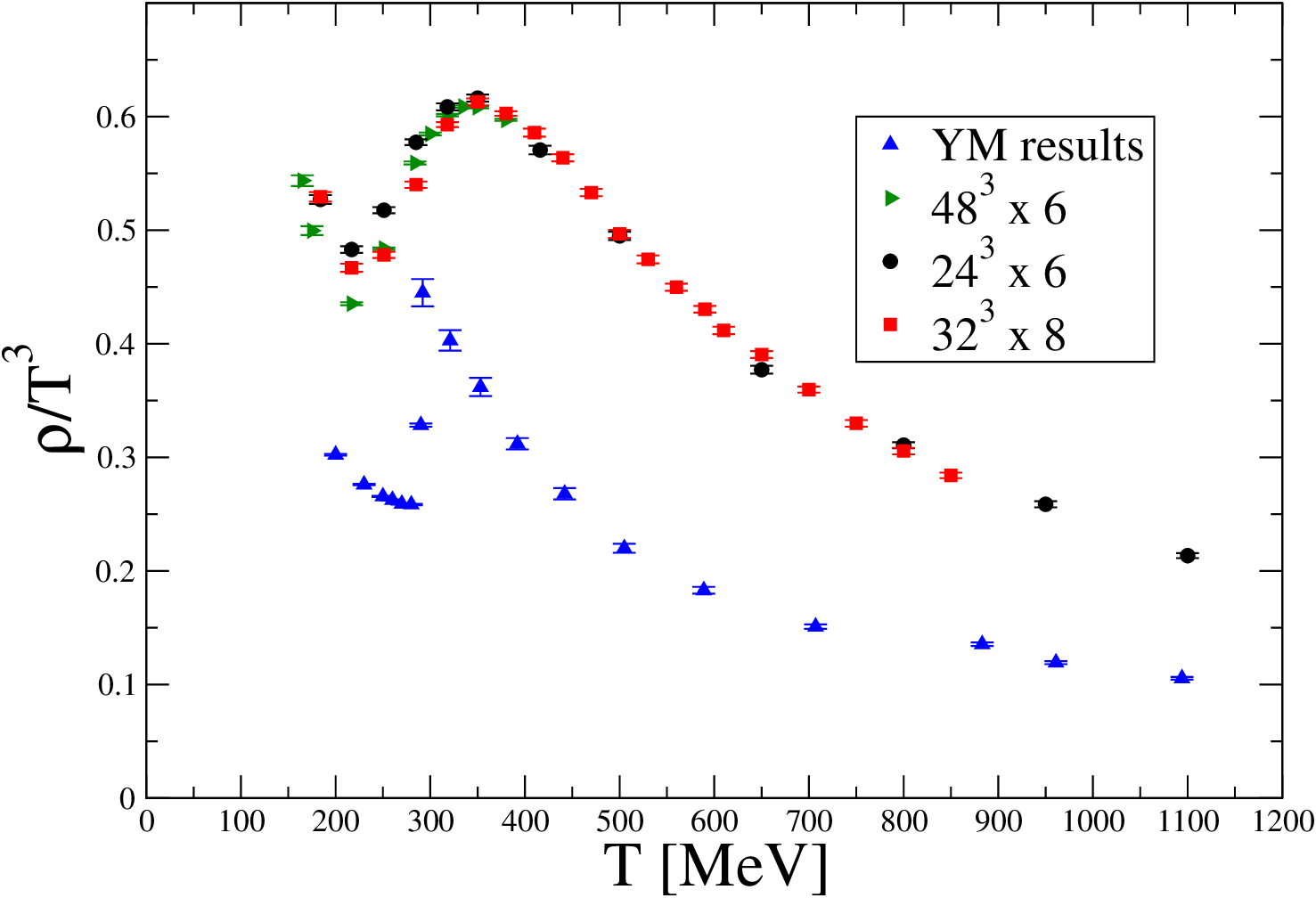}
\end{center}
\caption{Normalized thermal monopole density as a function 
of $T$, for full QCD and for the pure gauge theory.}
\label{fig4}
\end{figure}

The similarities with the pure gauge theory appear also when considering the 
normalized
total thermal monopole density $\rho$, which is reported
in Fig.~\ref{fig4}. At high $T$, $\rho/T^3$ in full QCD is about twice than
in pure gauge, but follows a similar behavior; in particular, data 
for $T \gtrsim 600$~MeV are reasonably fitted by the perturbative 
prediction~\cite{Liao:2006ry, Giovannangeli:2001bh}
\begin{equation}
{\rho}/{T^3} \propto (\log (T/\Lambda_{eff}))^{-3}\, 
\label{fitlog}
\end{equation}
with $\Lambda_{eff} = 47(5)$~MeV ($48(1)$~MeV for pure gauge~\cite{Bonati:2013bga}). 
Around the transition, pure gauge results show a sharp drop, which can be interpreted
as disappearance of part of the thermal component due to condensation: in practice, 
part of the thermal wrappings disappear because now monopole currents wrap also
along spatial directions. A similar behavior, even if smoother, is observed
for full QCD data in correspondence of $T_{BEC}$, resembling also 
the prediction of Ref.~\cite{Ramamurti:2017zjn} based on pressure data.

{\it Discussion} -- 
Standing as it is, and given that the interpretation of color confinement 
in terms of dual superconductivity and abelian monopole condensation is still not firmly 
assessed, our result cannot be considered as evidence for the existence of an 
intermediate phase of strongly interacting matter, where chirally symmetry is restored
but confinement is still at work. However, notwithstanding 
the ambiguities related to the Abelian projection procedure, there is no doubt 
that for pure gauge theories thermal monopoles catch many non-perturbative 
properties related to confinement.
It is therefore interesting to ask how our findings match with 
results from other standard observables.

Generally speaking, the non-perturbative region above the QCD crossover is quite large. Quark number susceptibilities
and thermodynamical quantities depart from those of a hadron gas starting 
from the chiral pseudocritical temperature $T_c \simeq 155$~MeV, however values 
compatible with a non-interacting Quark-Gluon Plasma are reached
only for $T \gtrsim 300$~MeV~\cite{Bazavov:2011nk,Bazavov:2014yba}.
The study of color screening properties confirms this picture, showing that 
in-medium quark-antiquark systems behave consistenty with weak-coupling picture 
only for $T \gtrsim 300$~MeV~\cite{Bazavov:2018wmo, Bazavov:2020teh}, which is the same 
temperature at which also violations to the Casimir scaling of the Polyakov loop 
disappear~\cite{Petreczky:2015yta}.

Another interesting hint comes from the study of $\theta$-dependence. 
It is well known that, in pure gauge theories, the topological charge distribution 
becomes compatible with that expected within the dilute instanton gas approximation
(DIGA) soon after $T_c$: this is visible especially from the kurtosis
coefficient $b_2$, which is compatible with the DIGA value, $b_2 = -1/12$, already
for $T \gtrsim 1.1~T_c$~\cite{b2_su3}. In full QCD, instead, $b_2$ approaches the DIGA 
value quite slowly, showing appreciable deviations still for $T > 2~T_c$~\cite{b2_qcd}. 
This has been interpreted in Ref.~\cite{Shuryak:2017fkh} in terms of the existence 
of an intermediate phase dominated by an instanton-dyons ensemble.

Finally, one should consider the recently advocated existence of intermediate
new phases of QCD, either the so-called stringy fluid 
phase~\cite{Glozman:2019fku} where strongly interacting matter 
would be chirally symmetric but still confined, 
supported numerically 
by an emergent enhanced symmetry in spatial meson 
correlators~\cite{Rohrhofer:2019qwq}, or the 
intermediate phase before the transition to a low-energy 
scale invariant regime, 
supported by 
numerical evidence on the properties of the lowest-lying part 
of the Dirac spectrum~\cite{Alexandru:2019gdm, Alexandru:2021pap}.

To summarize, there is already plenty of evidence 
for the region 
above the QCD crossover 
being dominated by peculiar non-perturbative effects: 
the detection of monopole condensation 
may help locating the transition to weak coupling 
more precisely. Therefore, our findings should be further 
investigated in various directions. First of all, one should clarify, by a careful
finite size scaling analysis, whether the 
observed condensation temperature $T_{BEC}$ 
corresponds to a real phase transition, even if
not associated with evident thermodynamical signatures. 
Then, since $T_{BEC}$ turns out to be strikingly close to the 
pure gauge $T_c$, one should investigate
whether this is accidental or not and what is the role 
of quarks in the game, by repeating our study for different 
quark masses and number of flavors, 
and by studying monopole-flavor correlations~\cite{Liao:2012tw}.
Finally, different confinement mechanisms
should be investigated, to check whether a similar transition temperature 
is detected independently, including also different order parameters within
the dual superconductor scenario, like the expectation value 
of magnetically charged operators~\cite{DelDebbio:1995yf, DiGiacomo:1999yas, DiGiacomo:1999fb, Carmona:2001ja, Carmona:2002ty}.


\noindent {\bf Acknowledgements:}
We thank Andrea Rucci for collaboration in the 
early stages of this study.
Numerical
simulations have been performed at the Scientific Computing Center at
INFN-PISA and on the MARCONI and M100 machines at CINECA, based on the agreement
between INFN and CINECA (under project INF19\_npqcd, INF20\_npqcd and INF21\_npqcd).

\end{document}